\documentclass{article}
\begin{document}
\title{
Modified Born-Jordan Method For Constructing The \\
Commutation Relation Of Coordinate and Momentum
\author{Ze Sen Yang  \\
Institute of Theoretical Physics, Peking University, Beijing 100871,
CHINA } }
\date{\today}
\maketitle
\begin{abstract}
      The Born-Jordan method for constructing the quantum condition
of the Matrix Mechanics is pointed out to be inappropriate in the
present work. We modify this method and reconstruct the quantum
condition by setting up a new expression for the Bohr quantum
condition with the help of the $(n,n)$ elements of the matrix
$\oint\hat{p}(t){\rm d}\hat{x}(t)$.
\end{abstract}
\vspace{1.0cm}
\begin{center}
{\bf I}.\ \ Introduction
\end{center} \par \ \par
\setcounter{equation}{0}
    It is well known that the quantum condition expressed as the
commutation relation of coordinate and momentum was first
constructed in the Matrix Mechanics by M.Born and P.Jordan in paper
${[1]}$. On the other hand the Born-Jordan method is inappropriate
and needs certain modifications. Owing to the fact that the
Born-Jordan method is a generlization of Heisenberg's method for
constructing the Heisenberg Quantum Condition we seek to reveal some
features of the drawback of the former method here by analyzing the
latter. Consider the special case that was studied by W.Heisenberg
in paper $[2]$. Denote by $x(t)$ and $m\,\dot{x}$ the coordinate and
momentum of periodic motion of one-dimensional system. The Bohr
"classical" quantum condition employed by Heisenberg can be written
as
\begin{eqnarray}
&& J_H = \int_0^{2\pi/\omega(n)} m\,\big(\dot{x}(n,t)\big)^2 {\rm d}
   t\,(= n\,h) =
    2\pi\,m\,\sum_{\alpha=-\infty}^{\infty}\alpha\,\big(\alpha\,\omega(n)\big)
        X_{-\alpha}(n)\,X_\alpha(n)\,,
\end{eqnarray}
with
\begin{eqnarray}
&&  x(n,t)= \sum_{\alpha=-\infty}^{\infty}X_\alpha(n)\,{\rm e}^
            {{\rm i}\,\omega(n)\,\alpha\,t}\,,
\end{eqnarray}
where the integer $n$ stands for the stationary states. According to
Heisenberg's rule $X_\alpha(n)$ and $\alpha\,\omega(n)/(2\pi)$
should be replaced by the transition amplitude $X_{n,n-\alpha}$
between states $(n,n-\alpha)$ and the transition frequency
$\omega(n,n-\alpha)/(2\pi)$. Moreover $X_{n,n+\alpha}$ is regarded
as $X^*_{n,n-\alpha}$ in paper $[2]$ so that $x(n,t)$ is a real
function. The formula $(1)$ thus becomes
\begin{eqnarray}
&&  J_H = 2\pi\,m\,\sum_{\alpha=-\infty}^{\infty}\alpha\,
             \omega(n,n-\alpha)X_{n,n+\alpha}\,X_{n,n-\alpha}\,,\nonumber\\
&& \hspace{0.6cm}
        = 2\pi\,m\,\sum_{\alpha=-\infty}^{\infty}\alpha\,
             \omega(n,n-\alpha)X^*_{n-\alpha,n}\,X_{n,n-\alpha}\,.
\end{eqnarray}
Next Heisenberg  replaced equation $1=\frac{\partial J_H}{\partial
J_H}$ with the difference equation
\begin{eqnarray*}
&& h=2\pi\,m\,\sum_{\alpha=-\infty}^{\infty}\,\alpha\,
\frac{\delta}{\delta n}\,\Big(
  \omega(n,n-\alpha)X^*_{n,n-\alpha}\,X_{n,n-\alpha} \Big)\,,
\end{eqnarray*}
and obtained his Quantum Condition
\begin{eqnarray}
&& \hbar= m\,\sum^{\infty}_{\alpha=-\infty}\big\{X^*_{n+\alpha,n}X_{n+\alpha,n}\,\omega(n+\alpha,n)\nonumber\\
&& \hspace{2.0cm} -
X^*_{n-\alpha,n}X_{n-\alpha,n}\,\omega(n,n-\alpha)\big\}\,,
\end{eqnarray}
which can be written as
\begin{eqnarray}
   {\rm i}\hbar =\big( X^\dagger\,P-P\,X^\dagger \big)_{nn}\,\,,
\end{eqnarray}
where the $(n,n-\alpha)$ elements of the matrices $X$ and $P$ are
$X_{n,n-\alpha}$ and $P_{n,n-\alpha}$ respectively and
$$
   P_{n,n-\alpha}= {\rm i}\,m\,\omega(n,n-\alpha)\,X_{n,n-\alpha}\,.
$$
However M.Born regarded $X^\dagger$ in the formula $(5)$ to be $X$.
Actually it is not allowed to do so because the condition
$X_{n,n+\alpha}=$ $X^*_{n,n-\alpha}$ together with the hermitian
property of $X$ leads to
\begin{eqnarray}
&& X_{n,n +\alpha}= X_{n-\alpha,n}\,,\ \ \
    X_{n+\alpha,n}=X_{n,n-\alpha}\,.
\end{eqnarray}
Such equations are incorrect and yield wrong results such as
$X_{n,n-1}=$ $X_{1,0}$ and  $X_{n-1,n}=$ $X_{0,1}$. It might be
worthwhile to see what happens when the formula $(4)$ together with
$X=X^\dagger$ is applied to the linear harmonic oscillator problem
with the Hamiltonian
$$
   H = \frac{1}{2m}\,P^2 + \frac{1}{2}m\,\omega^2\,X^2 \,.
$$
Thus
\begin{eqnarray*}
&& x_n(t)= X_{n,n-1}\,{\rm e}^{{\rm i}\,\omega\,t} + X_{n,n+1}\,{\rm e}^{-{\rm i}\,\omega\,t}\,,\\
&& \omega(n,n\pm1)=\mp\,\omega\,,\ \ \
   \omega(n\pm1,n)=\pm\,\omega\,.
\end{eqnarray*}
One therefore has from $(4)$ that
$$
  \hbar = 2m\,\omega \Big(X^*_{n+1,n}X_{n+1,n}-X^*_{n-1,n}X_{n-1,n} \Big)\,.
$$
If the matrix $X$ is hermitian then $X_{n,n\pm 1}$ is independent of
$n$ and the right hand side of this formula is equal to zero.
\par
    What can be learned is that the formula $(1)$ and $(3)$ for
the Bohr quantum condition is incompatible with the hermitian
property of the matrices $X$ and $P$ and should not be employed to
construct the quantum condition of the Matrix Mechanics.
\par\ \par
\vspace{0.5cm}
\begin{center}
{\bf II}.\ \ The Drawback of the Born-Jordan Method
\end{center} \par \ \par
    For carefully examining the drawback of the Born-Jordan method we now analyze the
steps of constructing the quantum condition in paper $[1]$. Let
$x(t)$ and $p(t)$ be the canonical coordinate and momentum of
periodic motion of one-dimensional system. In the first step, M.Born
and P.Jordan followed Heisenberg's paper $[2]$ to express Bohr's
"classical" quantum condition. Their formula can be written as$\,:$
\begin{eqnarray}
&& J_{BJ}= \oint p(n,t){\rm d}x_(n,t)= \int_0^{2\pi/\omega(n)}
     p(n,t)\dot{x}(n,t){\rm d}t = n\,h + J_0 \nonumber\\
&& \hspace{0.70cm}
     = 2\pi\,m\,\sum_{\alpha=-\infty}^{\infty}\alpha\,P_{-\alpha}(n)\,X_\alpha(n)\,,
\end{eqnarray}
with
\begin{eqnarray}
&&  x(n,t)= \sum_{\alpha=-\infty}^{\infty}X_\alpha(n)\,{\rm e}^{{\rm i}\,\omega(n)\,\alpha\,t}\,,\\
&&  p(n,t)= \sum_{\alpha=-\infty}^{\infty}P_\alpha(n)\, {\rm
e}^{{\rm i}\,\omega(n)\,\alpha\,t}\,,
\end{eqnarray}
Now the underline assumption here is that the coordinate and
momentum and their real functions are represented by hermitian
matrices, the $(n,n')$ elements of $\hat{x}(t)$ and $\hat{p}(t)$ are
as follows
$$
 x_{nn'}(t)= X_{n,n'}\,{\rm e}^{{\rm i}\,\omega(n,n')\,t} \,, \ \ \
 p_{nn'}(t)= P_{n,n'}\,{\rm e}^{{\rm i}\,\omega(n,n')\,t} \,.
$$
After replacing  $X_\alpha(n)$ and $P_\alpha(n)$ by $X_{n,n-\alpha}$
and  $X_{n,n-\alpha}$ the formula $(7)$ becomes
\begin{eqnarray}
&& J_{BJ}= 2\pi\,{\rm i}\sum_{\alpha=-\infty}^{\infty}\alpha\,
           P_{n,n+\alpha}X_{n,n-\alpha}
  =-2\pi\,{\rm i}\sum_{\alpha=-\infty}^{\infty}\alpha\,P_{n,n-\alpha}X_{n,n+\alpha} \,.
\end{eqnarray}
   Step\,(2)\,,\,The equation $1=\frac{\partial J_{BJ}}{\partial J_{BJ}}$ was replaced
with the difference equation
\begin{eqnarray}
&& h=2\pi\,{\rm i}\,\sum_{\alpha=-\infty}^{\infty}\,\alpha
\frac{\delta}{\delta n}\,\Big(P_{n,n +\alpha} \,
        X_{n,n-\alpha}\Big)\,,
\end{eqnarray}
which was interpreted as
\begin{eqnarray}
&& {\rm i}\,\hbar =\sum_{\alpha=-\infty}^{\infty}
        \big(P_{n-\alpha,n}X_{n,n-\alpha}-P_{n,n+\alpha}X_{n+\alpha,n}\big)\,\,,
\end{eqnarray}
or
\begin{eqnarray}
&& {\rm i}\hbar=\big( X\,P-P\,X\big)_{nn}\,\,.
\end{eqnarray}
For the case $p(t)=$$m\,\dot{x}(t)$, substituting $P_{n-\alpha,n}=$
${\rm i}m\,\omega(n-\alpha,n) X_{n-\alpha,n}$ and $P_{n,n+\alpha}=$
$P^*_{n+\alpha,n}$ in equation $(12)$ leads to
\begin{eqnarray}
&& \hbar= m\,\sum^{\infty}_{\alpha=-\infty}\big\{X_{n,n+\alpha}X_{n+\alpha,n}\,\omega(n+\alpha,n)\nonumber\\
&& \hspace{2.0cm} -
X_{n,n-\alpha}X_{n-\alpha,n}\,\omega(n,n-\alpha)\big\}\,,
\end{eqnarray}
which was regarded as Heisenberg's quantum condition or Thomas-Kuhn
equation$^{[3,4]}$.
\par
    Step\,(3)\,,\,With the help of the equation of motion M.Born and P.Jordan came to the conclusion that the
matrix $\hat{x}(t)\hat{p}(t)-\hat{p}(t)\hat{x}(t)$ is independent of
time and is diagonal.
\par
    The first two steps sought to carry through and extend Heisenberg's
method of paper $[2]$ under the condition that matrices $X$ and $P$
are hermitian. The matter is not so simple However. Obviously,
further assumptions should be introduced for getting the formula
$(12)$ from $(11)$. It seems that the assumption $X_{n,n+\alpha}=$
$X_{n-\alpha,n}$ or $P_{n,n+\alpha}=$ $P_{n-\alpha,n}$ is useful for
going from equation $(11)$ to $(12)$ but they are unreasonable
themselves. As stated in Sec.I $X_{n,n+\alpha}=$ $X_{n-\alpha,n}$,
which together with $X=$ $X^\dagger$ means $x(n,t)=$ $x^*(n,t)$,
leads to wrong results and therefore makes the formula $(12)$
invalid. Similarly. $P_{n,n+\alpha}=$ $P_{n-\alpha,n}$, which
together with $P=$ $P^\dagger$ means $P(n,t)=$ $P^*(n,t)$, also
yield $P_{n,n+1}=$ $P_{0,1}$ and other wrong results and makes the
formula $(12)$ invalid. In particular, if both assumptions
$x_{n,n+\alpha}=$ $X_{n-\alpha,n}$ and $P_{n,n+\alpha}=$
$P_{n-\alpha,n}$ are employed then the right hand side of the
formula $(12)$ equates to zero.
\par
    Actually, it can be verified that the Born-Jordan method is not available for constructing the
quantum condition of the Matrix Mechanics. For the special case p(t)
$p(t)=$$m\,\dot{x}(t)$, the formula $(10)$ becomes
\begin{eqnarray}
&&  J_{BJ} = 2\pi\,m\,\sum_{\alpha=-\infty}^{\infty}\alpha\,
             \omega(n,n-\alpha)X_{n,n+\alpha}\,X_{n,n-\alpha}\,,
\end{eqnarray}
which should not be regarded as Heisenberg's formula $(3)$ because
$X$ is hermitian. Let us now suppose the formula $(14)$ can be
obtained reasonably from $(15)$ and consider again the linear
harmonic oscillator problem one has
\begin{eqnarray}
&& \hbar =
      2\,m\,\omega\,\Big\{X_{n+1,n}X_{n,n+1} - X_{n,n-1}X_{n-1,n}\Big\}\,.
\end{eqnarray}
From this formula and other principles of the Matrix Mechanics the
coordinate momentum commutation relation can be established (see the
end of Sect.3) and the elements of the Matrix $X$ can also be found.
On the other hand the simplified form of the formula $(15)$ is
\begin{eqnarray*}
&&  J_{BJ} = 2\pi\,m\,\sum_{\alpha=\pm 1}
         \alpha\,\omega(n,n-\alpha)\,X_{n,n-\alpha}X_{n,n+\alpha}\,.
\end{eqnarray*}
Consequently the Quantum Condition can also be written as
\begin{eqnarray*}
&& \hbar = m\,\omega\,\sum_{\alpha=\pm 1}
        \alpha\,\Big\{X_{n+\alpha,n}X_{n+\alpha,n+2\alpha} - X_{n,n-\alpha}X_{n,n+\alpha}\Big\}\\
&& \hspace{1.0cm}= m\,\omega\,\Big\{X_{n+1,n}X_{n+1,n+2}
                            - X_{n-1,n}X_{n-1,n-2}\Big\}\,.
\end{eqnarray*}
It can easily be checked that this formula is invalid.
\par
    To sum up, with the Born-Jordan method it is not possible to
construct the quantum condition reasonably even for linear harmonic
oscillator problem.
\par
\vspace{0.5cm}
\begin{center}
{\bf III}.\ \  Modification of the Born-Jordan Method and \\
               the Reconstruction of the Quantum Condition
\end{center} \par \ \par
    In order to modify the Born-Jordan Method and reconstruct
the quantum condition, it is natural to express the Bohr quantum
condition with reference to the $(n,n)$ element of the matrix
$\oint\hat{p}(t){\rm d}\hat{x}(t)$. Since
\begin{eqnarray}
&& \Big(\hat{p}(t){\rm d}\hat{x}(t)\Big)(n,n)
     = - \sum_\alpha P_{n,n-\alpha}\,X_{n-\alpha,n}\,{\rm i}\,\omega(n,n-\alpha){\rm d}t\,,\\
&& \Big(\hat{p}(t){\rm d}\hat{x}(t)\Big)^*(n,n)
     = \sum_\alpha P_{n-\alpha,n}\,X_{n,n-\alpha}\,
       {\rm i}\,\omega(n,n-\alpha){\rm d}t\,, \\
&& \int^{2\pi/\omega(n)}_0 p(n,t)\,\dot{x}^*(n,t){\rm d}t
      = -2\pi\,{\rm i}\sum_\alpha \alpha\,P_\alpha(n)\,X^*_\alpha(n)\,,\nonumber\\
&& \hspace{2.30cm} = -2\pi\,{\rm i}\sum_\alpha
                       \alpha\,P_{n,n-\alpha}\,X^*_{n,n-\alpha}\,,
\end{eqnarray}
and $\omega(n,n-\alpha)$ corresponds to the classical quantities
$\alpha\,\omega(n)$, one sees that $\oint\big(\hat{p}(t){\rm d}
\hat{x}(t)\big)(n,n)$ corresponds to $\oint p(n,t)\, {\rm
d}x^*(n,t)$. It is thus reasonable to write the Bohr quantum
condition as$\,:$
\begin{eqnarray}
  J=\frac{1}{2}\Big\{\oint p^*(n,t)\,{\rm d}x(n,t) + \oint p(n,t)\,{\rm d}x^*(n,t)\Big\}=n\,h + J_0 \,,
\end{eqnarray}
It is obvious from $(17)$ that $\frac{\delta}{\delta n}\,\oint
p(n,t)\,{\rm d}x^*(nt)$ is real, namely
\begin{eqnarray}
&& \frac{\delta}{\delta n}\oint p(n,t)\,{\rm d}x^*(n,t)
   =\frac{\delta}{\delta n}\oint p^*(n,t)\,{\rm d}x(n,t)
   \nonumber\\
&&\hspace{1.0cm} = -2\pi\,{\rm i}\sum_\alpha
   P_{n+\alpha,n}X_{n,n+\alpha} + 2\pi\,{\rm i}\sum_\alpha P_{n,n-\alpha}X_{n-\alpha,n}\,.
\end{eqnarray}
Therefore $1=\frac{\partial J}{\partial J}$ is equivalent to
$h=\frac{\partial }{\partial n}\oint p(n,t)\,{\rm d}x^*(n,t)$ and
the Quantum Condition is obtained from the difference equation
\begin{eqnarray}
&& h = \frac{\delta}{\delta n}\oint p(n,t)\,{\rm d}x^*(n,t)\,,
\end{eqnarray}
or
\begin{eqnarray}
&& {\rm i}\hbar =\sum_{\alpha=-\infty}^{\infty}
        \big(P_{n+\alpha,n}X_{n,n+\alpha}-P_{n,n+\alpha}X_{n+\alpha,n}\big)\nonumber\\
&& \hspace{2.0cm}=\big( X\,P-P\,X\big)_{nn}\,\,.
\end{eqnarray}
\par
    For the case  $p(t)=$$m\,\dot{x}(t)$, the above new expression for the Bohr quantum condition becomes
\begin{eqnarray}
&&  J =n\,h+J_0= \oint p^*(n,t){\rm d}x(n,t)= \oint p(n,t){\rm d}x^*(n,t)\nonumber\\
&&\hspace{2.02cm}
      = 2m\pi\,\sum_{\alpha=-\infty}^{\infty} \alpha\,\omega(n,n-\alpha)X_{n,n-\alpha}X_{n-\alpha,n}\,.
\end{eqnarray}
It follows from the formula $(22)$ or $(23)$ that
\begin{eqnarray}
&& \hbar= m\,\sum^{\infty}_{\alpha=-\infty}\big\{X_{n,n+\alpha}X_{n+\alpha,n}\,\omega(n+\alpha,n)\nonumber\\
&& \hspace{2.0cm}
      - X_{n,n-\alpha}X_{n-\alpha,n}\,\omega(n,n-\alpha)\big\}\,.
\end{eqnarray}
It should be remembered that owing to $X_{n\pm\alpha,n}=$
$X^*_{n,n\pm\alpha}\neq$ $X_{n,n\mp\alpha}$ this formula is not
equivalent to Heisenberg's quantum condition.
\par
    Finally it is an easy task to establish the commutation relation
$\big(X\,P-P\,X\big)=$ ${\rm i} \hbar\,$ from it's diagonal part by
improving Born-Jordan's argument (see also the footnote (58) of
reference $[5]$). In their paper $[1]$ M.Born and P.Jordan judged
the conservation matrix $\hat{x}(t)\hat{p}(t)-\hat{p}(t)\hat{x}(t)$
to be diagonal by employing an assumption that $\omega(n,n')\neq 0$
for $n\neq n'$. However B.L.Van Der Waerden pointed out in reference
$[6]$ that since Born-Jordan's argument relied on this assumption
they did not give an exact mathematical proof of the formula for
$\big( X\,P-P\,X\big)$. Actually, the whole independent stationary
states $\{n\}$ can be defined to form a representation so that both
of the conservation quantity
$\big(\hat{x}\hat{p}-\hat{p}\hat{x}\big)/({\rm i}\hbar)$ and the
Hamiltonian are diagonal. Consequently one sees from the formula
$(23)$ that $\big(X\,P-P\,X\big)/({\rm i}\hbar)$ is an unit matrix.
\par
\vspace{0.5cm}
\begin{center}
{\bf IV}.\ \ Concluding Remarks
\end{center} \par \ \par
    Historically (see Waerden's paper of reference
$[6]$), M.Born rewrote Heisenberg's Quantum Condition as ${\rm i}
\hbar=$ $(X\,P-P\,X)_{nn}$ and guessed the matrix $(X\,P-P\,X)$ to
be diagonal. One of the main tasks of Born-Jordan's paper $[1]$ is
to verify M.Born's guess. Correspondingly, in the present work we
first verify that Heisenberg's Quantum Condition can not be
rewritten in Born's way and therefore the formula $(3)$ should not
be applied to the Matrix Mechanics. We then show that Born-Jordan
method, which is based on the formulas $(7)$, can not be employed to
construct the quantum condition of the Matrix Mechanics. It is also
clear from our argument that even if Heisenberg's assumption
$X_{n,n-\alpha}=$ $X^*_{n,n+\alpha}$ is excluded, his formula $(3)$
could not be applied to the Matrix Mechanics.
\par
    Furthermore a new formula is setting up to express the Bohr quantum condition
with the help of the $(n,n)$ elements of the matrix
$\oint\hat{p}(t){\rm d}\hat{x}(t)$. This new formula is employed to
modify the Born-Jordan method and reconstruct the quantum condition.
\par
    The author is grateful to Professor Zhao Kai Hua and Professor Cheng Tan Sheng for helpful discussions.
\par
\ \par
\begin{center}
{\large \bf References}
\end{center}
\par  \noindent
[1]\ M.Born and P.Jordan, Z.Phys. {\bf 34}, 858-888(1925).
\par  \noindent
[2]\ W.Heisenberg, Z.Phys. {\bf 33}, 879-893(1925).
\par  \noindent
[3]\ W.Thomas, Naturwiss. {\bf 13}, 627(1925).
\par  \noindent
[4]\ W.Kuhn, Z.Phys. {\bf 33}, 408-412(1925).
\par  \noindent
[5]\ W.A.Fedak and J.J.Prentis, Am.J.Phys. {\bf 77}, 128-139(2009).
\par  \noindent
[6]\ B.L.van der Waerden, Sources of Quantum Mechanics, Dover,
      New York, 1968.
\end{document}